\documentclass[11pt,a4paper]{article}
\usepackage{jcappub} % for details on the use of the package, please
\usepackage[T1]{fontenc} % if needed
\usepackage{graphicx,epsfig}% Include figure files
\usepackage{epstopdf}

\title{\boldmath Constraining the range of Yukawa gravity interaction from S2
star orbits III: improvement expectations for graviton mass bounds}

\author[a,b,c,d,e,1]{A. F. Zakharov,\note{Corresponding author.}}
\author[f]{P. Jovanovi\'{c},}
\author[g]{D. Borka}
\author[g]{and V. Borka Jovanovi\'{c}}

\affiliation[a]{National Astronomical Observatories of Chinese
Academy of Sciences,\\
Datun Road 20A, Beijing, 100012 China}
\affiliation[b]{Institute of Theoretical and Experimental Physics,
117259 Moscow, Russia}
\affiliation[c]{National Research Nuclear University MEPhI \\
(Moscow Engineering Physics Institute), 115409, Moscow, Russia}
\affiliation[d]{Bogoliubov Laboratory for Theoretical Physics, JINR,
141980 Dubna, Russia}
\affiliation[e]{North Carolina Central University, Durham, NC 27707,
USA}
\affiliation[f]{Astronomical Observatory, Volgina 7, P.O. Box
74, 11060 Belgrade, Serbia}
\affiliation[g]{Atomic Physics Laboratory (040), Vin\v{c}a
Institute of Nuclear Sciences, \\
University of Belgrade, P.O. Box 522, 11001 Belgrade, Serbia}

\emailAdd{zakharov@itep.ru} \emailAdd{pjovanovic@aob.rs}
\emailAdd{dusborka@vinca.rs} \emailAdd{vborka@vinca.rs}

\abstract{Recently, the LIGO-Virgo collaboration discovered
gravitational waves and in their first publication on the subject
the authors also presented a graviton mass constraint as $m_g <
1.2 \times 10^{-22}$~eV \cite{Abbott_16} (see also more details in
a complimentary paper \cite{Abbott_16_tests}). In our previous
papers we considered constraints on Yukawa gravity parameters
\cite{Borka_13} and on graviton mass from analysis of the trajectory
of S2 star near the Galactic Center \cite{Zakharov_16}. In the paper
we analyze a potential to reduce  upper bounds for graviton mass
with future observational data on trajectories of bright stars near
the Galactic Center. Since gravitational potentials are different
for these two cases, expressions for relativistic advance for
general relativity and Yukawa potential are different functions
on eccentricity and semimajor axis, it gives an opportunity to
improve current estimates of graviton mass with future observational
facilities. In our considerations of an improvement potential for a
graviton mass estimate we adopt a conservative strategy and assume
that trajectories of bright stars and their apocenter  advance will be
described with general relativity expressions and it gives
opportunities to improve graviton mass constraints. In contrast with
our previous studies, where we present current constraints on
parameters of Yukawa gravity \cite{Borka_13} and graviton mass
\cite{Zakharov_16} from observations of S2 star, in the paper we
express expectations to improve current constraints for graviton
mass, assuming the GR predictions about apocenter shifts will be
confirmed with future observations. We concluded that if future
observations of bright star orbits during around fifty years will
confirm GR predictions about apocenter shifts of bright star orbits
it give an opportunity to constrain a graviton mass at a level
around $5 \times 10^{-23}$~eV or slightly better than current
estimates obtained with LIGO observations.
}


\begin{document}
\maketitle
\flushbottom
\section{Introduction}

\subsection{Theories of massive gravity}

A development of general relativity for more than 100 years was
extremely successful. Currently, predictions of GR have been
confirmed with many different experiments and observations. Recently,
gravitational waves and binary black holes have been discovered with
LIGO detectors. Now, one can say that "Einstein was 100\% right"
\cite{Damour_06} but the correctness of the claim may be changed with
new observational and experimental data and GR phenomena must be
checked again and again with better precision. There are many
proposals to check GR predictions with future observations and
experiments. However, if we adopt GR as the universal theory of
gravity, then we face the dark matter (DM) and dark energy (DE)
problems. There is a slow progress to find solutions of the DM and DE
puzzles. There exists a natural way to avoid DM and DE problems with
changing the fundamental gravity law. Now there are many alternative
theories of gravity and some of these theories have no Newtonian
limit in a weak gravitational field approximation. A theory of
massive gravity where graviton has non-vanishing mass is among
popular alternative theories of gravity. A direct graviton detection
is an extremely difficult problem \cite{Dyson_13}, however, if a
graviton has a mass one can discuss opportunities to constrain its
mass with current and future observations.  In the paper we discuss
the issue in more details. A theory of massive gravity was
introduced by M. Fierz and W. Pauli in 1939 \cite{Fierz_39}. At the
beginning a development of the theory was very slow. In seventies of
the last century a couple of pathologies of such a gravity theory
have been found, such as van Dam -- Veltman -- Zakharov --
Iwasaki discontinuity \cite{vanDam_70,Zakharov_70,Iwasaki_70} for
$m_g \rightarrow 0$ (where $m_g$ is a graviton mass) and if one
considers a deflection of light, for instance, in the framework of
Fierz -- Pauli theory with vanishing graviton mass, one obtains a
result which is different from the GR expression. However, the
analysis presented in these papers based on linear approximation of
weak gravitational field and as it was shown in \cite{Vainshtein_72}
the approximation is valid only for $r \gg r_V$ (where $r_V$ is the
Vainshtein radius and $r_V = \left(\dfrac{M}{m_g}\right)^{1/5}$ in
Planck units, therefore, for typical stellar masses and current
graviton mass estimates the length is much greater than a size of
Solar system). The phenomenon is called now Vainshtein screening
(see, also
\cite{Deffayet_2002,Rubakov_08,Babichev_09,Babichev_09_2,Zee_10}
for a more detailed discussion).

Boulware and Deser found a presence of ghosts and instabilities in
theory of massive gravity \cite{Boulware_72,Boulware_72b}. Nowadays,
a number of different techniques have been proposed to construct
theories of massive gravity without Boulware -- Deser ghosts
\cite{Rubakov_08,Hinterbichler_12}.
A class of ghost-free massive gravity has been proposed in papers
\cite{deRham_10,deRham_11}, such a theory are called now  de Rham --
Gabadadze -- Tolley (dRGT) gravity model (see also more comprehensive
reviews \cite{deRham_14,deRham_17}).

A. A. Logunov and his group proposed a bi-metric theory of gravity
with a massive graviton and investigated its cosmological
consequences in papers
\cite{Logunov_88,Chugreev_89,Gershtein_03,Gershtein_04, Logunov_06,
Gershtein_07} at the period of low popularity of such theories of
massive gravity due to a presence of pathologies such as
discontinuities and ghosts, see also recent publications of this group and references therein \cite{Soloviev_17,Soloviev_17_TMP,Chugreev_17}.

In the last years it has been a great progress in theoretical
development of massive gravity theories and they may be treated now
as a reasonable alternative for the conventional GR and people
discuss observational signatures of such theories.

\subsection{Constraints on parameters of Yukawa gravity}

%\textbf{
Massive gravity  is a specific case of Yukawa gravity.
Any universal theoretical model of gravity has to be checked at
different scales: in laboratory, Solar system \cite{Talmadge_88}, our Galaxy
(including both its innermost region around the central supermassive
black hole, as well as its outer parts), other galaxies,
clusters of galaxies and at cosmological scales.
Geophysical,  astronomical and laboratory constraints on Yukawa
gravity are summarized in Fig. 9 (for 1~$\mu$m $\le \Lambda \le$
1~cm), and Fig. 10 (for 1~cm $\le \Lambda \le 10^{15}$m) in
 \cite{Adelberger_09}. From these data one can see that there is a
growing trend for strength of Yukawa interaction with increase of
its range, especially for $\Lambda > 10^{11}$ m. Also, one can see
that gravity is relatively well constrained at short ranges (see
Figs. 9 and 10 in \cite{Adelberger_09}). Solar System, Lunar Laser
Ranging (LLR) and LAGEOS constraints give $\Lambda \gg 1.5 \times 10^{11}$ m
and $\Lambda \gg 4 \times 10^8$ m \cite{Adelberger_09,Moffat_08}.
Clearly, that forthcoming data analysis of LARES data will improve LAGEOS constraints
on Yukawa theory parameters \cite{Ciufolini_13,Ciufolini_13b,Ciufolini_16}.
 Toy models with
the Yukawa potential derived in $f(R)$ gravity have been considered for Solar System and the Galactic Center
and it was shown how to evaluate Yukawa potential parameters from observations for these systems
 \cite{Demartino_18,De Laurentis_18}.  Yukawa potential  have been successfully used to fit observational data for galactic rotation curves and
 clusters
of galaxies \cite{Cardone_11,Capozziello_07,Capozziello_09}.
It is important to note that there are not too many
 constraints for parsec (and sub-parsec) scales and such constraints could be obtained
from an analysis of star trajectories in the
Galactic Central Parsec and we try to evaluate expectations to improve
previous results on graviton mass constraints with S2 star trajectory. The obtained constraints from S-star
orbits ($10^{15} \le \Lambda \le 10^{18}$ m) are beyond the parameter
region given in Figs. 9 and 10 from Ref. \cite{Adelberger_09}. Besides, it
is very important to investigate gravity in the vicinity of black holes
to probe a gravity law in a strong gravitational field limit.
%}

\subsection{Observational constraints on graviton mass}

Probably first estimates of graviton mass have been given by F.
Zwicky  as $m_g < 5 \times 10^{-64}$g \cite{Zwicky_61} (to obtain
this estimate he assumed that the Newtonian law has to be valid until
the typical scales for galactic clusters\footnote{Hiida and
Yamaguchi used a similar assumption in paper \cite{Hiida_65} where
they obtained the estimate $m_g < 5 \times 10^{-62}$g. As V. L.
Ginzburg noted a couple of times, in astronomy ten is equal to one,
moreover, in astronomy estimates for lengths and masses (and related
quantities) are significantly changing with time because an
evaluation of lengths in astronomy is often model dependent.} (some
time before Zwicky discussed an opportunity that the fundamental
gravity law has to be valid for scales of galactic clusters because
in this case it is checked in a reliable way while for longer length scales it may
be changed \cite{Zwicky_57} and for such scales an impact of the
cosmological $\Lambda$-term  may be significant). M. Hare used
Galactic scale to estimate Compton length for graviton \cite{Hare_73}
$m_g < 1.2 \times 10^{-66}$g. If one uses the cosmological length
scale \cite{Gershtein_03,Gershtein_04}, then $m_g < m_H^0$, where
$m_H^0= \dfrac{\hbar H_0}{c^2}=3.8 h \times 10^{-66}$~g is "Hubble
mass" and $H_0= h_{100} \times$ 100 km/(s Mpc) is the current Hubble
constant ($h_{100}$ is a useful dimensionless parameter).

Observations of pulsars give an opportunity to evaluate a graviton
mass. As it is well-known binary pulsars provide a remarkable test
of GR predictions that their orbits have to be shrinking due to
gravitational radiation and it firstly was observed for Hulse --
Taylor pulsar PSR B1913+16 \cite{Hulse_75} (see also
\cite{Weisberg_16} for a more modern review). In paper \cite{Finn_02}
Finn and Sutton showed that  evolutions of orbits for the binary
pulsars PSR B1913+16 and PSR B1534+12 constrained  a graviton mass
at a level $7.6 \times 10^{-20}$~eV with 90\% C.L.
A review of opportunities to evaluate parameters of alternative theories of gravity with
binary pulsars is given \cite{Esposito_11}.
In paper \cite{Jones_05}
it was evaluated an opportunity to estimate a graviton mass from future LISA observations of
evolution of eccentric binary systems.

As it was noted many years ago predicted and observed time of
arrivals for pulsar residuals may be used for detection of
gravitational waves with long wavelengths \cite{Sazhin_78}. This idea
is widely used to detect (or constrain) long gravitational waves with
pulsar timing observations \cite{Desvignes_16,Lam_17}.
As it was shown in \cite{Lee_10}, current and
future pulsar timing arrays could be used not only for detection of
gravitational waves but also to obtain  graviton mass constraints for
different programs for observations at a level around
$[3 \times 10^{-23}, 3 \times 10^{-22}]$~eV as and later Lee showed
that these constraints can be improved with a more sophisticated data
analysis \cite{Lee_13}.

In paper \cite{Brito_13} the authors considered study of massive spin-2 fluctuations (including massive gravitons) around
Schwarzschild and slowly rotating Kerr black holes and concluded that such black holes
are  generically unstable and solid observations of such objects will lead to conclusion that graviton mass is
less than $5 \times 10^{-23}$~eV.

Considering weak lensing for galactic clusters, one could obtain a more stringent constraint on a graviton mass
$6 \times 10^{-32}$~eV \cite{Choudhury_04}, in the paper the authors argued that the corresponding Compton wavelength
is more than 100~Mpc, while an analysis of CMB anisotropy leads to conclusion that Compton wavelength should be more than
4~Mpc \cite{Binetruy_01}.

In papers \cite{Fomalont_03,Kopeikin_04} a minimal speed of gravity
$v_g=c(1-0.15)$ has been evaluated from the relativistic light
deflection of the quasar J0842+1835 by gravitational field of Jupiter
on September 8, 2002. In principle one could evaluate a graviton
mass from the constraint on speed of gravity and the dispersion
relation.

A number of other ways to constrain a graviton mass from
observations are given in comprehensive reviews
\cite{deRham_17,Goldhaber_10}. One should note that very often when
people discussed observational constraints on graviton mass they
presented their expectations or forecasts from future observations
since uncertainties and systematics were not carefully analyzed.
Therefore, such estimates are model dependent.

\subsection{Observational constraints on graviton mass from
observations of gravitational waves}

In times when theories of massive gravity were not very popular due
a presence of pathologies C. Will considered an opportunity to
evaluate a graviton mass from observations of gravitational waves
\cite{Will_98} (see also \cite{Will_14} for a more detailed
discussion).

Assuming that a graviton mass is small in comparison with energy of
gravitational waves $h f \gg m_g c^2$, then
\begin{equation}
v_g/c \approx 1-
\dfrac{1}{2}\left(\dfrac{c}{\lambda_g f}\right)^2,
\label{velocity}
\end{equation}
where
$\lambda_g=h/(m_g c)$ is the graviton Compton wavelength and one
could obtain
\cite{Will_98}
\begin{equation}
\lambda_g > 3 \times 10^{12} \mathrm{km} \left(\frac{D}{200~\mathrm{Mpc}} \frac{100~\mathrm {Hz}}{f}\right)^{1/2}\left(\frac{1}{f\Delta t} \right)^{1/2}, \label{Dispersion_Eq 4}
\end{equation}
\begin{equation}
\Delta t = \Delta t_a - (1+z)\Delta t_e, \label{Dispersion_Eq 3}
\end{equation}
where $\Delta t_a = t^{EM}_a - t^{GW}_a$, $\Delta t_e = t^{EM}_e -
t^{GW}_e$, $t^{EM}_a (t^{EM}_e)$ and  $t^{GW}_a (t^{GW}_e)$ are
arrival (emission) instant of electromagnetic radiation and arrival
(emission) instant for gravitational waves. As it was pointed out
\cite{Will_98}, one can use Eq. \eqref{Dispersion_Eq 4} if observers
detected gravitational waves and electromagnetic radiations from one
source and $\Delta t_e$ is known or can be evaluated with a
sufficient accuracy. Moreover, there is an opportunity to constrain
a graviton mass in the case if there is only a gravitational wave
signal. For numerical estimate, one can estimate $f \Delta t \sim
\rho^{-1} \approx 10$ (where $\rho$ is a signal-to-noise ratio) for
LIGO-Virgo ground based interferometers \cite{Will_98}, therefore,  a
graviton mass constraint can be at a level $2.5 \times 10^{-22}$~eV
for ground based LIGO-Virgo detectors.

The LIGO-Virgo collaboration reported about the first detection of
gravitational waves from a merger of two black holes (it was detected
on September 14, 2015 and it is called GW150914) \cite{Abbott_16}.
According to estimates from the shape of gravitational wave signal
the source is located at a luminosity distance of around 410~Mpc
(which corresponds to a redshift $z \approx 0.09$), the initial
black hole masses were $36 M_\odot$ and $29 M_\odot$ and the final
black hole mass is $62 M_\odot$, therefore, around $3 M_\odot$ was
emitted in gravitational waves in 0.1 s. The collaboration not only
discovered gravitational waves but also detected the first binary
black hole system and one of the most powerful source of radiation in
the Universe and the energy was released in gravitational waves.
Moreover, the team constrained the graviton Compton wavelength
$\lambda_g > 10^{13}$~km which could be interpreted as a constraint
for a graviton mass $m_g < 1.2 \times 10^{-22}$~eV \cite{Abbott_16}
(the estimate roughly coincides with theoretical predictions
\cite{Will_98}).

In June 2017 the LIGO-Virgo collaboration published a paper where
the authors described a detection of gravitational wave signal from a
merger of binary black hole system with masses of components
$31.2M_\odot$ and $19.4 M_\odot$ at distance around 880~Mpc which
corresponds to $z \approx 0.18$ \cite{Abbott_17a}. In this case,
around $2M_\odot$ was emitted in gravitational waves in around 0.4~s.
The event is named GW170104. In this paper the authors significantly
improved their previous graviton mass constraint, $m_g  < 7.7 \times
10^{-23}$~eV \cite{Abbott_17a}.

On August 17, 2017 the LIGO-Virgo collaboration detected a
merger of binary neutron stars with masses around $0.86 M_\odot$ and
$2.26 M_\odot$ at a distance around 40 Mpc (GW170817) and after 1.7 s
the Fermi-GBM detected $\gamma$-ray burst GRB 170817A associated with
the GW170817 \cite{Abbott_17b,Abbott_17c}. Since gravitational wave
signal was observed before GRB 170817A one could conclude that the
observational data are consistent with massless or very light
graviton, otherwise, electromagnetic signal could be detected before
gravitational one because in the case of relatively heavy gravitons
gravitational waves could propagate slower than light.

In the consideration one assumes that photon is massless (but graviton may be massive).
In the case of massive photon     $m_ \gamma>0$ (see, for instance \cite{Jackson_98} for introduction of Proca theory which describes  a massive photon case)
to use the same logic   at least we have to have $(c-v_\gamma) << (c - v_g)$ ($c$ is a limiting speed of ultra high energy quanta,
 $v_\gamma$ and $v_g$ are velocities of quanta and gravitons respectively) or
\begin{equation}
m_g/f >> m_{\gamma}/\nu,
\label{photon_mass}
\end{equation}
as we see from Eq. (\ref{velocity}), where $m_g$ and $m_\gamma$ are masses of graviton and photon, respectively; $f$ and $\nu$ are their typical frequencies) and photon mass is constrained with another experimental (or observational) data.
Different ways to evaluate photon mass are discussed in couple of reviews \cite{Goldhaber_10,Tu_05,Okun_06} and original papers \cite{Luo_03,Tu_06,Ryutov_07,Wu_16,Bonetti_16,Bonetti_17}. Laboratory experiments gave the upper limit as
$m_{\gamma}<7\times 10^{-19}$~eV \cite{Luo_03} or $m_{\gamma}<5 \times 10^{-20}$~eV \cite{Tu_06}, while astrophysical constraint from analysis of plasma
in Solar wind gave $m_{\gamma}<10^{-18}$~eV \cite{Ryutov_07}, analysis of Fast Radio Bursts gave weaker constraints on photon mass $m_{\gamma}<10^{-14}$~eV
\cite{Wu_16,Bonetti_16,Bonetti_17}.

One could roughly estimate frequency band for quanta where inequality (\ref{photon_mass}) is hold.
If we adopt the upper limit of graviton mass (around $ 10^{-22}$~eV) obtained by the LIGO collaboration from the first GW events without electromagnetic counterpart and we assume $f \approx 100$, then the inequality (\ref{photon_mass}) is hold for spectral band of quanta from radio up to higher frequencies if we use upper limit estimates from papers
\cite{Luo_03,Tu_06,Ryutov_07} and the inequality (\ref{photon_mass}) is hold for spectral band of quanta from optical band up to higher frequencies
if we use upper limit estimates from papers
\cite{Wu_16,Bonetti_16,Bonetti_17}.

Constraints on
speed of gravitational waves have been found
$-3\times 10^{-15} < (v_{g}-c)/c < 7 \times 10^{-16}$   \cite{Abbott_17c}. Graviton
energy is $E=hf$, therefore, assuming a typical LIGO frequency range
$f \in (10,100)$, from the dispersion relation one could obtain a
graviton mass estimate $m_g < 3 \times (10^{-21}-10^{-20})$~eV which
a slightly weaker estimate than previous ones obtained from binary
black hole signals detected by the LIGO team
\cite{Zakharov_IHEP_2017}.

\subsection{Observations of bright stars near the Galactic Center}

Two groups of observers are monitoring bright stars (including S2
one) to evaluate gravitational potential at the Galactic Center
\cite{Ghez_00,Rubilar_01,Schodel_02,Ghez_03,Ghez_04,Ghez_08,
Gillessen_09a,Gillessen_12,Meyer_12,Zhang_15,Yu_16,Gillessen_17,
Hees_17,Hees_17_Moriond,Chu_17}. The precision observations of S2
star \cite{Gillessen_09b} were used to evaluate parameters of black hole
and to test and constrain several models of modified gravity at mpc
scales \cite{Nucita_07,Zakharov_07,Borka_12,Zakharov_14,Zakharov_15,Zakharov_16}.
The simulations of the S2 star orbit around the supermassive black
hole at the Galactic Centre (GC) in Yukawa gravity \cite{Borka_13}
and their comparisons with the NTT/VLT astrometric observations of
S2 star \cite{Gillessen_09b} resulted with the constraints on the
range of Yukawa interaction $\Lambda$. These constrains showed that
$\Lambda$ is most likely on the order of several thousand
astronomical units. Assuming Yukawa gravitational potential of a
form $\propto r^{-1}\exp(-r/\lambda_g)$ \citep[see e.g.][]{Will_98}
this result indicates that it can be used to constrain the lower
bound for Compton wavelength $\lambda_g$ of the graviton, i.e. the
upper bound for its mass
\begin{equation}
m_{g(upper)}=h\,c/\lambda_g.
\label{mass_vs_lambda}
\end{equation}

\section{Estimates}

The goal of this paper is to discuss perspectives to reduce an upper
limit for graviton mass $m_{g}$ with the future observations of
bright stars assuming the bright star apocenter shifts calculated in
the framework of classical GR will be confirmed with future
observations, therefore, analyzing possible range for
$\Lambda_{\mathrm crit}$ parameters, we conclude that $\Lambda$
parameters corresponding to greater shifts $\Lambda
<\Lambda_{\mathrm crit}$ should be ruled out with these
observations. We think that our assumption is rather conservative
because earlier GR predictions were always confirmed (if the GR
predictions will not be confirmed with future observations we have
to introduce new object such as a bulk distribution of stellar
cluster or dark matter or to change a fundamental gravity law
according to Le Verrier's suggestions \cite{Zakharov_09}).
Therefore, we have the condition for $\Lambda_{\mathrm crit}$, that
Yukawa gravity induces the same orbital precession as GR. Assuming
that $\Lambda_{\mathrm crit}$ gives a lower limit for Compton length
for graviton one obtains an upper graviton mass constraint. Thus, we
expect to improve a current estimate of graviton mass with the
future observations analyzing apocenter advances in Yukawa potential
and GR.

In other words, we consider $\Lambda$ parameter of Yukawa gravity
which induces the same orbital precession as predicted by GR. We
also show that, if such precessions will be detected in the observed
orbits of S-stars around the Galactic Center, they could be used to
put more stringent constraints on the range of Yukawa interaction, as
well as on the mass of graviton. For that purpose, we first find the
conditions which must be satisfied by the parameters of Yukawa
gravity in order to induce the same stellar orbits as GR, then we
use these forecasts to constrain the range of Yukawa interaction
and graviton mass, taking into account the latest results about
orbital parameters of stellar orbits at the Galactic Center (GC),
given in \cite{Gillessen_17,Hees_17}. We also perform graphical
comparisons of the simulated stellar orbits with the same apocenter
shifts in these two theories of gravity.

\subsection{Orbital precession due to general central-force
perturbations}

A general expression for apocenter shifts for Newtonian potential and
small perturbing potential is given as a solution of problem 3 in
Section 15 in the classical Landau \& Lifshitz (L \& L) textbook
\cite{Landau_76}.\footnote{See applications of the expressions for
calculations of stellar orbit precessions in presence of supermassive
black hole and dark matter \cite{Dokuchaev_15,Dokuchaev_15a}.}
In paper \cite{Adkins_07}, the authors derived the expression which
is equivalent to the (L \& L) relation in an alternative way and
showed that the expressions are equivalent  and after that they
calculated apocenter shifts for several examples of perturbing
functions.

According to \cite{Adkins_07}, orbital precession $\Delta\varphi$
per orbital period, induced by small perturbations to the Newtonian
gravitational potential $\Phi_N(r)=-\dfrac{GM}{r}$ could be evaluated
as:
\begin{equation}
\Delta\varphi^{rad} = \dfrac{-2L}{GM e^2}\int\limits_{-1}^1
{\dfrac{z
\cdot
dz}{\sqrt{1 - z^2}}\dfrac{dV\left( z \right)}{dz}},
\label{2.1}
\end{equation}
while in the textbook \cite{Landau_76}
it was given in the form
\begin{equation}
\Delta\varphi^{rad} = \dfrac{-2L}{GM e}\int\limits_{0}^\pi
\cos \varphi r^2 \dfrac{dV( r )}{dr} d \varphi,
\label{2.2}
\end{equation}
\noindent where $V(z)$ is the perturbing potential, $r$ is related
to $z$ via:
$r = \dfrac{L}{1 + ez}$ in Eq. (\ref{2.1}) (and $r = \dfrac{L}{1 + e
\cos \varphi}$ in Eq. (\ref{2.2})) , and $L$ being the semilatus
rectum of the orbital ellipse with semi-major axis $a$ and
eccentricity $e$:

\begin{equation}
L = a\left( {1 - {e^2}} \right),
\label{semilatus}
\end{equation}

\noindent while $\Delta\varphi$ represents true precession in the
orbital plane, and the corresponding apparent value $\Delta s$, as
seen from Earth at distance $R_0$, is \cite{Weinberg_05} (assuming
that stellar orbit is perpendicular to line of sight and taking into
account an inclination of orbit one has to write an additional
factor which is slightly  less than 1 in the following expression):

\begin{equation}
\Delta s\approx\dfrac{a(1+e)}{R_0}\Delta\varphi .
\label{prec_apparent}
\end{equation}

%Similar results we can find in paper \cite{ursu17}.

In order to compare the orbital precession of S-stars in both GR
and Yukawa gravity, we applied the same procedure as described in
\cite{Borka_13} to perform the two-body simulations of the stellar
orbits in the framework of these two theories.

\subsection{Orbital precession in General Relativity}

In order to simulate the orbits of S-stars in GR we used
parameterized post-Newtonian (PPN) equation for acceleration of a
test particle orbiting the central mass, given in \cite{Anderson_75}.
From their Eqs. (11) and (12), one can easily obtain the following
PPN equation of motion for two-body problem \cite{Anderson_75}:

\begin{equation}
\vec{\ddot{r}}=-G (M+m)\dfrac{\vec{r}}{r^3}+\dfrac{G
M}{c^2r^3}\left\lbrace\left[2\left(\beta+\gamma\right)\dfrac{G
M}{r}-\gamma\left(\vec{\dot{r}}\cdot\vec{\dot{r}}\right)\right]\vec{r
}
+2\left(1+\gamma\right)\left(\vec{r}\cdot\vec{\dot{r}}\right)\vec{
\dot{r}}
\right\rbrace.
\label{ppn}
\end{equation}

\noindent The second term in r.h.s. of the above expression
(\ref{ppn}) represents the PPN correction to the acceleration due to
contribution of central mass, and  it depends on two PPN parameters:
$\beta$ and $\gamma$. In the case of GR: $\beta=1$ and $\gamma=1$,
and then this PPN correction induces the well known expression for
Schwarzschild precession \citep[see e.g.][]{Will_14,Weinberg_72}:

\begin{equation}
\Delta\varphi^{rad}\approx\dfrac{2\pi G M}{c^2
a(1-e^2)}\left(2+2\gamma-\beta\right)\quad\Rightarrow\quad
\Delta\varphi_{GR}^{rad}\approx\dfrac{6\pi G M}{c^2 a(1-e^2)}.
\label{gr_prec}
\end{equation}

The corresponding formula for apparent precession $\Delta s_{GR}$,
as seen from Earth at distance $R_0$, could be calculated
according to \citep{Weinberg_05}:

\begin{equation}
\Delta s_{GR}\approx\dfrac{6\pi G M}{c^2 (1-e)R_0} ,
\label{gr_prec_apparent}
\end{equation}

\noindent which in the case of GR does not depend on $a$.

\subsection{Orbital precession in Yukawa gravity}

In order to simulate orbits of S-stars in Yukawa gravity we assumed
the following gravitational potential \citep[see e.g.][and references
therein]{Borka_13}:

\begin{equation}
\Phi_Y(r)=-\dfrac{GM}{(1+\delta)r}\left[{1+\delta
e^{-\dfrac{r}{\Lambda}}}
\right],
\label{yukawa_pot}
\end{equation}

\noindent where $\Lambda$ is the range of Yukawa interaction
and $\delta$ is a universal constant. Here we will assume that
$\delta > 0$, as indicated by data analysis of astronomical
observations \citep[see e.g.][]{Cardone_11,Capozziello_09}. Yukawa
gravity induces a perturbation to the Newtonian gravitational
potential described by the following perturbing potential:

\begin{equation}
V_Y(r)=\Phi_Y(r)+\dfrac{GM}{r}=\dfrac{\delta}{1+\delta}\dfrac{GM}{r}
\left[{1-
e^{-\dfrac{r}{\Lambda}}}
\right]
\label{yukawa_pert}
\end{equation}

\noindent The exact analytical expression for orbital precession in
the case of the above perturbing potential could be presented in the
integral form Eqs. (\ref{2.1}) and (\ref{2.2}), but we will
calculate the approximate expression for $\Delta\varphi$ using power
series expansion of $V_Y(r)$, assuming that $r\ll\Lambda$:

\begin{equation}
V_Y(r)\approx-\dfrac{\delta G M
r}{2(1+\delta)\Lambda^2}\left[1-\dfrac{r}{3\Lambda}+\dfrac{r^2}{
12\Lambda^2}-\ldots \right],\quad r\ll\Lambda,
\label{yukawa_pert_approx}
\end{equation}

\noindent where we neglected the constant term since it does not
affect $\Delta\varphi$. By substituting the above expression into
(\ref{2.1}) we obtain the following approximation for the angle of
orbital precession in Yukawa gravity:

\begin{equation}
\Delta\varphi_Y^{rad}\approx\dfrac{\pi\delta\sqrt{1-e^2}}{1+\delta}
\left(\dfrac{a^2}{\Lambda^2}-\dfrac{a^3}{\Lambda^3}+\dfrac{4+e^2}{8}
\dfrac{a^4}{
\Lambda^4 }-\ldots\right).
\label{yukawa_prec_series}
\end{equation}

\noindent
As it was shown in \cite{Adkins_07} the right-hand side in Eq.
(\ref{yukawa_prec_series}) could be presented as series of Gauss's
hypergeometric function $_2F_1$ with different arguments.

Since $r\ll\Lambda$ also implies that $a\ll\Lambda$, we can neglect
higher order terms in the above expansion. The first order term then
yields the following approximate formula for orbital precession:
\begin{equation}
\Delta\varphi_Y^{rad}\approx\dfrac{\pi\delta\sqrt{1-e^2}}{1+\delta}
\dfrac{a^2}{
\Lambda^2},\quad a\ll\Lambda.
\label{yukawa_prec}
\end{equation}
\noindent Both, $\Delta\varphi_{GR}$ and $\Delta\varphi_Y$
represent the angles of orbital precession per orbital period in the
orbital plane (i.e. true precession). The corresponding apparent
values in Yukawa case $\Delta s_{Y}$, as seen from Earth at distance
$R_0$ is (for $\delta =1$), according to paper \citep{Weinberg_05} :

\begin{equation}
\Delta s_{Y}\approx\dfrac{a(1+e)}{R_0}\Delta\varphi_Y^{rad}\approx
0.5\pi  \frac{a^3}{R_0\Lambda^2} (1+e)\sqrt{1-e^2}.
\label{prec_apparent_Y}
\end{equation}
If one believes that a gravitational field at the Galactic Center is
described with a Yukawa potential, then the maximal $\Delta
s_{Y}$ value corresponds to $e=1/2$ when function
$(1+e)\sqrt{1-e^2}$ has its maximal value (assuming that all other
parameters are fixed).

\subsection{Expectations to constrain the range of Yukawa gravity
with future observations}

We assume that in future GR predictions about precession angles for
bright star orbits around the Galactic Center will be successfully
confirmed, therefore, for each star we have a constraint on
$\Lambda$ which can be obtained from the condition for $\Lambda$, so
that Yukawa gravity induces the same orbital precession as GR. This
constraint can be obtained directly from (\ref{gr_prec}) and
(\ref{yukawa_prec}), assuming that
$\Delta\varphi_{Y}=\Delta\varphi_{GR}$. In this way we obtain
that:

\begin{equation}
\Lambda\approx\sqrt{\dfrac{\delta c^2 (a\sqrt{1-e^2})^3}{6
(1+\delta)
G M}}.
\label{lambda}
\end{equation}

\noindent As it can be seen from the above expression, taking into
account that $\delta$ is universal constant, the corresponding values
of $\Lambda$ in the case of all S-stars depend only on the semi-major
axes and eccentricities of their orbits. In order to stay in
accordance with \cite{Zakharov_16}, here we will also assume that
$\delta=1$, in which case formula (\ref{lambda})
reduces to:

\begin{equation}
\Lambda\approx\dfrac{c}{2}\sqrt{\dfrac{(a\sqrt{1-e^2})^3}{3 G M}}\approx \sqrt{\dfrac{(a \sqrt{1-e^2})^3}{6R_S}},
\label{lambda_delta1}
\end{equation}

Using Kepler law  we could write the previous equation in the
following form
\begin{equation}
\Lambda\approx \dfrac{T}{T_0}\sqrt{\dfrac{(a_0 \sqrt{1-e^2})^3}{6R_S}}.
\label{lambda_delta1_T}
\end{equation}
where $T_0$ and $a_0$ are periods and semi-major axis for a selected
orbit of S2 star, for instance. Inspecting Eq.
(\ref{lambda_delta1_T}) one can see that greater constraints on
$\Lambda$ parameter (smaller graviton mass bounds) correspond to
orbits with longer periods and smaller eccentricities (usually orbits
with small eccentricities are not interesting to test GR precession
phenomena because shifts for more eccentric orbits are greater).

\section{Numerical results}

\begin{figure*}[ht!]
\centering
\includegraphics[width=0.5\textwidth]{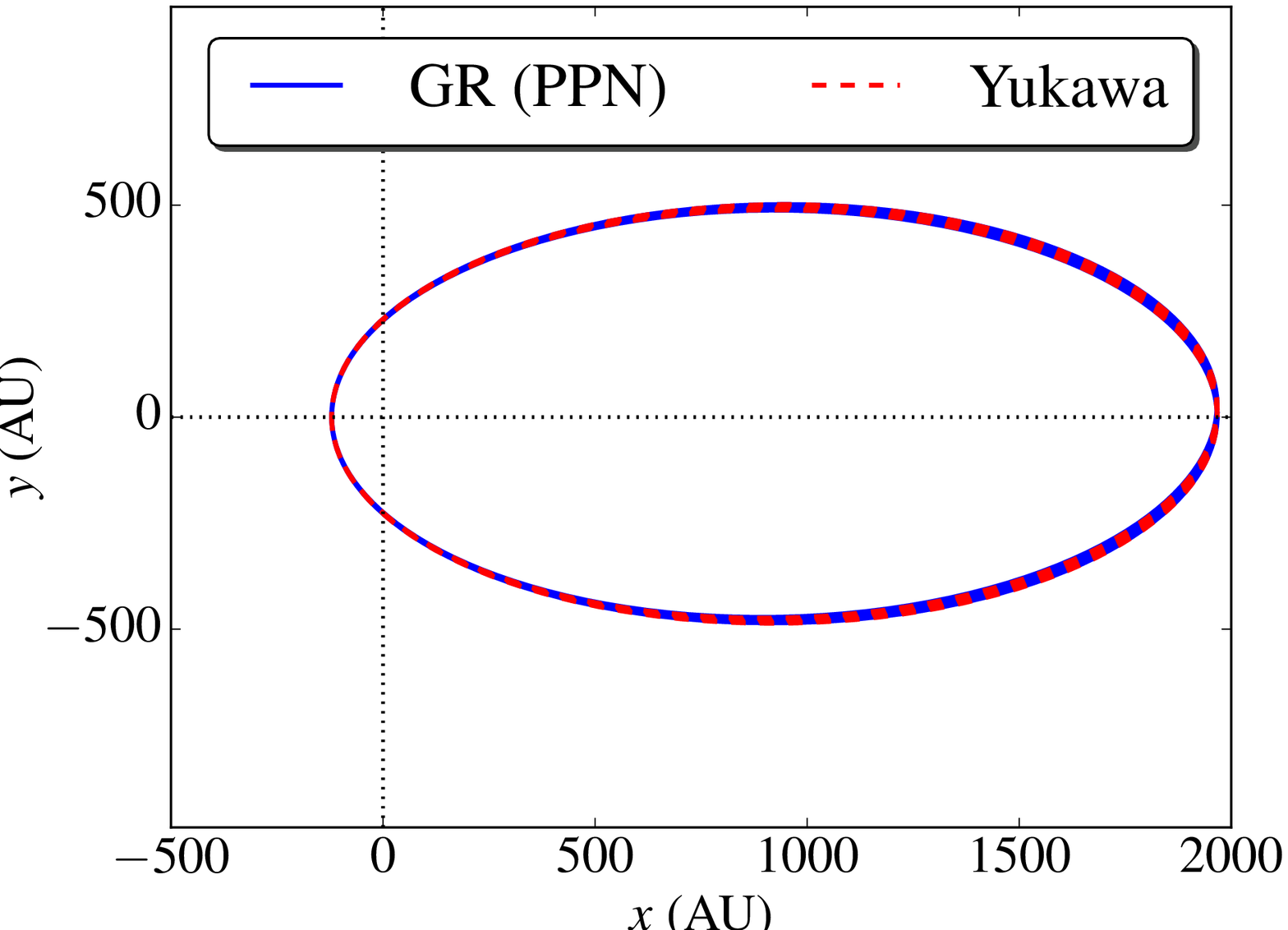}\hfill
\includegraphics[width=0.5\textwidth]{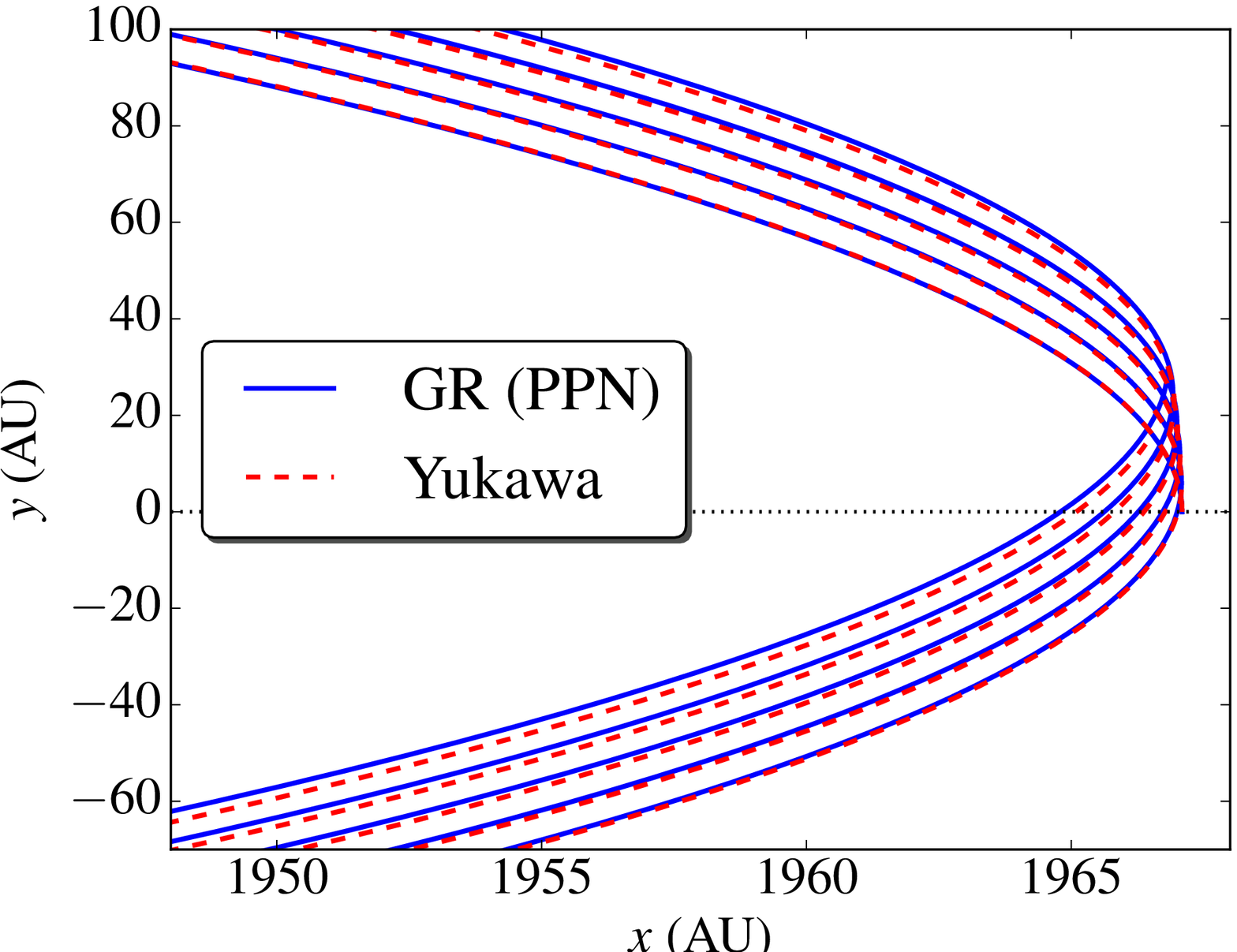}
\caption{Comparison between the simulated orbits of S2-star in GR
(blue solid line) and in Yukawa gravity (red dashed line) during five
orbital periods. Region around apocenter is zoomed in the right panel
in order that small orbital precession of $\Delta\varphi=722.''1=
0.^\circ 2$ is visible.}
\label{fig01}
\end{figure*}

We first applied the formula (\ref{lambda_delta1}) to estimate the
value of $\Lambda$ in the case of S2-star, so that its orbits in GR
and Yukawa gravity almost coincide with each other. For that purpose
we used the results presented in \cite{Gillessen_17}, according to which
mass of the central black hole of the Milky Way is $M=4.28\times
10^6\ M_\odot$, distance to the GC is $R_0=8.32\ \mathrm{kpc}$,
semi-major axis of the S2-star orbit is $a=0.''1255$, and its
eccentricity $e = 0.8839$. The corresponding range of Yukawa gravity
according to Eq. (\ref{lambda_delta1}) is $\Lambda = 15125.5\
\mathrm{AU}$. We then compared the orbital precession of S2-star in
both GR and Yukawa gravity by performing the two-body simulations of
its orbit in the same way as described in \cite{Borka_13}. Fig.
\ref{fig01} presents a graphical comparison between the obtained
simulated orbits during five orbital periods. As it can be seen from
this figure, the values of parameter $\Lambda$ calculated from
(\ref{lambda}) produce the simulated orbits in Yukawa gravity which
are almost identical to those in GR, therefore, if the geodesics for
these two different theories of gravity produce the same apocenter
advances, they have tiny differences in their shapes.

Besides, we also used this approach to calculate the values
for $\Lambda$ in the case of all S-stars from the Table 3 of
\cite{Gillessen_17} except S111. The observed orbital
elements, together with their uncertainties, of these stars are
given in Table \ref{tab1}, while the obtained numerical results are
presented in Table \ref{tab2}, where $\Delta\varphi$ and $\Delta s$
represent orbital precession in both GR and Yukawa gravity, since
their numerical values are identical up to three decimals, at least.

The obtained estimates of $\Lambda$ are then used to find the
corresponding constraints for the graviton mass $m_g$, according to
Eq. (\ref{mass_vs_lambda}),
where we assumed that the Compton wavelength $\lambda_g$
of graviton is equal to the obtained values of $\Lambda$:
$\lambda_g=\Lambda$. The resulting graviton masses are also given in
Table \ref{tab1}. The first column in Table \ref{tab1} contains the
observed semi-major axes of all S-stars which are converted from
arcsec to AU using the following distance to Sgr A$^\ast$: $R_0=8.32\
\mathrm{kpc}$  \cite{Gillessen_17}.

The relative errors of the range of Yukawa interaction and
graviton mass can be obtained by differentiating the logarithmic
versions of  Eqs. (\ref{lambda_delta1},\ref{mass_vs_lambda}), which gives the
following expression:
\begin{equation}
\dfrac{\Delta\Lambda}{\Lambda}=\dfrac{\Delta
m_g}{m_g}\approx\pm\dfrac{3}{2}\left(\dfrac{\left|\Delta
a\right|}{a}+\dfrac{e\left|\Delta e\right|}{1-e^2}
+\dfrac{1}{3}
\dfrac{\left|\Delta
M\right|}{M}
\right),
\label{relerr}
\end{equation}
Currently, we know that $ \Delta
M/{M}$ is around 7.2\% \cite{Gillessen_17} is a relatively small quantity
and it is the same for all orbits,
so that the uncertainty contributes  the relative error (around 3.6\%) to relative errors
which are presented in column 6 of Table \ref{tab2}. There is a hope that in future the relative error for the
black hole mass estimate will be improved.
We used the above analytic expression to estimate the numerical
values of relative and absolute errors for $\Lambda$ and $m_g$ which
are also given in Table \ref{tab2}. As it can be seen from this
table, the obtained relative errors range from a few
percent (as e.g. for S6 and S38) up to more than a hundred of percent
(as e.g. for S54 and S85), depending on the current observational
uncertainties for the semi-major axis and eccentricity of the S-star
orbits (see Table \ref{tab1}).  On the other
hand, here we showed that the relative error of the graviton mass
estimate in the case of the S2 star is $\approx 5.8\%$ (see Table
\ref{tab2}). Therefore, one could expect that with future observations
apocenter shift for S2 star one should obtain graviton mass estimate around $5.48 \times 10^{-22}$~eV with
relative error  $\approx 5.8\%$,
while, for instance, for S4 star with orbital period $T=77$~yr we have $4.1 \times 10^{-23}$~eV with
relative error  $\approx 5.6\%$,
assuming current uncertainties for mass, eccentricity and semi-major axis.
The expectation for graviton mass estimate with S4 star trajectory is better than the graviton mass LIGO estimate obtained from  GW170104 event.
One should note that a) current estimates for errors of orbital parameters and mass are changing with time since subsequent observations will
be performed; b) GR predictions about pericenter shifts will be confirmed with additional uncertainties; c) when subsequent observations will be done, perhaps, a theoretical model for motion of bright stars near the Galactic Center should be clarified, namely one should add into accout a bulk mass distribution of stellar cluster and dark matter. However, one should keep in mind that bulk distribution of matter introduces apocenter shifts
in the opposite direction in respect to relativistic advance \cite{Rubilar_01,Nucita_07,Zakharov_07} but Yukawa potential and PN potential in GR
cause apocenter advances. It gives an opportunity to compare values for these two shifts in positive directions.

As can be seen from Table \ref{tab1} for all studied S-stars
obtained results are $\gtrsim 10^{-24}\ \mathrm{eV}$, and it
indicates that this value represents a limit of graviton mass which
could be reached with the discussed procedure. Besides, one can see
that the stars with the largest semi-major axes and periods (e.g.
R44) give the smallest upper bounds on graviton mass (but orbital
periods are too long for these cases), while the stars with the
shortest orbits (e.g. S2) give the highest upper limits on graviton
mass. Therefore, more stringent constraints on graviton mass could be
achieved in future from observed orbits of S-stars with longer
periods and, preferably, lower eccentricities.

\clearpage
\begin{table*}
\centering
\caption{The observed orbital elements and their uncertainties for all S-stars in Table 3 from \cite{Gillessen_17},
except S111. The semi-major axes are converted from arcsec to AU assuming the following distance
to Sgr~A$^\ast$: $R_0 = 8.32$~kpc \cite{Gillessen_17}.}
\setlength{\tabcolsep}{0.11cm}
\begin{tabular}{|l|rcl|rcl|rcl|}
\hline
\multicolumn{1}{|c|}{Star}&\multicolumn{3}{c|}{$a\pm\Delta a\ \mathrm{(AU)}$}&\multicolumn{3}{c|}{$e\pm\Delta e$}&
\multicolumn{3}{c|}{$T\pm\Delta T\ \mathrm{(yr)}$} \\
\hline
\hline
S1   &  4950.4 &$\pm$& 199.7 & 0.556 &$\pm$& 0.018 & 166.0 &$\pm$& 5.8 \\
S2   &  1044.2 &$\pm$& 7.5 & 0.8839 &$\pm$& 0.0019 & 16.00 &$\pm$& 0.02 \\
S4   &  2970.2 &$\pm$& 30.8 & 0.3905 &$\pm$& 0.0059 & 77.0 &$\pm$& 1.0 \\
S6   &  5469.6 &$\pm$& 5.0 & 0.8400 &$\pm$& 0.0003 & 192.0 &$\pm$& 0.17 \\
S8   &  3367.1 &$\pm$& 11.6 & 0.8031 &$\pm$& 0.0075 & 92.9 &$\pm$& 0.41 \\
S9   &  2266.4 &$\pm$& 34.1 & 0.644 &$\pm$& 0.020 & 51.3 &$\pm$& 0.70 \\
S12  &  2485.2 &$\pm$& 15.0 & 0.8883 &$\pm$& 0.0017 & 58.9 &$\pm$& 0.22 \\
S13  &  2197.3 &$\pm$& 13.3 & 0.4250 &$\pm$& 0.0023 & 49.00 &$\pm$& 0.14 \\
S14  &  2382.0 &$\pm$& 30.0 & 0.9761 &$\pm$& 0.0037 & 55.3 &$\pm$& 0.48 \\
S17  &  2961.1 &$\pm$& 79.9 & 0.397 &$\pm$& 0.011 & 76.6 &$\pm$& 1.0 \\
S18  &  1979.3 &$\pm$& 12.5 & 0.471 &$\pm$& 0.012 & 41.9 &$\pm$& 0.18 \\
S19  &  4326.4 &$\pm$& 782.1 & 0.750 &$\pm$& 0.043 & 135 &$\pm$& 14 \\
S21  &  1822.1 &$\pm$& 14.1 & 0.764 &$\pm$& 0.014 & 37.00 &$\pm$& 0.28 \\
S22  & 10899.2 &$\pm$& 2329.6 & 0.449 &$\pm$& 0.088 & 540 &$\pm$& 63 \\
S23  &  2105.0 &$\pm$& 99.8 & 0.56 &$\pm$& 0.14 & 45.8 &$\pm$& 1.6 \\
S24  &  7854.1 &$\pm$& 399.4 & 0.8970 &$\pm$& 0.0049 & 331 &$\pm$& 16 \\
S29  &  3561.0 &$\pm$& 158.1 & 0.728 &$\pm$& 0.052 & 101.0 &$\pm$& 2.0 \\
S31  &  3735.7 &$\pm$& 83.2 & 0.5497 &$\pm$& 0.0025 & 108 &$\pm$& 1.2 \\
S33  &  5466.2 &$\pm$& 216.3 & 0.608 &$\pm$& 0.064 & 192.0 &$\pm$& 5.2 \\
S38  &  1178.1 &$\pm$& 1.7 & 0.8201 &$\pm$& 0.0007 & 19.2 &$\pm$& 0.02 \\
S39  &  3078.4 &$\pm$& 124.8 & 0.9236 &$\pm$& 0.0021 & 81.1 &$\pm$& 1.5 \\
S42  &  7904.0 &$\pm$& 1497.6 & 0.567 &$\pm$& 0.083 & 335 &$\pm$& 58 \\
S54  &  9984.0 &$\pm$& 7238.4 & 0.893 &$\pm$& 0.078 & 477 &$\pm$& 199 \\
S55  &   896.9 &$\pm$& 8.3 & 0.7209 &$\pm$& 0.0077 & 12.80 &$\pm$& 0.11 \\
S60  &  3225.7 &$\pm$& 58.2 & 0.7179 &$\pm$& 0.0051 & 87.1 &$\pm$& 1.4 \\
S66  & 12496.6 &$\pm$& 790.4 & 0.128 &$\pm$& 0.043 & 664 &$\pm$& 37 \\
S67  &  9368.3 &$\pm$& 216.3 & 0.293 &$\pm$& 0.057 & 431 &$\pm$& 10 \\
S71  &  8095.4 &$\pm$& 332.8 & 0.899 &$\pm$& 0.013 & 346 &$\pm$& 11 \\
S83  & 12396.8 &$\pm$& 1580.8 & 0.365 &$\pm$& 0.075 & 656 &$\pm$& 69 \\
S85  & 38272.0 &$\pm$& 27456.0 & 0.78 &$\pm$& 0.15 & 3580 &$\pm$& 2550 \\
S87  & 22796.8 &$\pm$& 1331.2 & 0.224 &$\pm$& 0.027 & 1640 &$\pm$& 105 \\
S89  &  8993.9 &$\pm$& 457.6 & 0.639 &$\pm$& 0.038 & 406 &$\pm$& 27 \\
S91  & 15949.4 &$\pm$& 740.5 & 0.303 &$\pm$& 0.034 & 958 &$\pm$& 50 \\
S96  & 12471.7 &$\pm$& 474.2 & 0.174 &$\pm$& 0.022 & 662 &$\pm$& 29 \\
S97  & 19302.4 &$\pm$& 3827.2 & 0.35 &$\pm$& 0.11 & 1270 &$\pm$& 309 \\
S145 &  9318.4 &$\pm$& 1497.6 & 0.50 &$\pm$& 0.25 & 426 &$\pm$& 71 \\
S175 &  3444.5 &$\pm$& 324.5 & 0.9867 &$\pm$& 0.0018 & 96.2 &$\pm$& 5.0 \\
R34  & 15059.2 &$\pm$& 1248.0 & 0.641 &$\pm$& 0.098 & 877 &$\pm$& 83 \\
R44  & 32448.0 &$\pm$& 11648.0 & 0.27 &$\pm$& 0.27 & 2730 &$\pm$& 1350 \\
\hline
\end{tabular}
\label{tab1}
\end{table*}

\begin{table*}
\centering
\caption{The estimated true and apparent orbital precession (columns 3 and 4), range of Yukawa interaction (column 5)
and graviton mass for $\lambda_g=\Lambda$ (column 6), as well as the corresponding absolute errors, for all S-stars
in Table 3 from \cite{Gillessen_17}, except S111. The last column contains the corresponding relative errors calculated
according to expression (\ref{relerr}), assuming the following mass of the central black hole: $M=4.28\pm 0.31\times 10^6\ M_\odot$ \cite{Gillessen_17}.
The orbital period obtained from the third Kepler's law is also shown in column 2.}
\setlength{\tabcolsep}{0.11cm}
\begin{tabular}{|l|r|r|c|rcl|rcl|r|}
\hline
\multicolumn{1}{|l|}{Star}&\multicolumn{1}{c|}{$T_{Kep}$}&\multicolumn{1}{c|}{$\Delta\varphi$}&\multicolumn{1}{c|}{$\Delta s$}&
\multicolumn{3}{c|}{$\Lambda\pm\Delta\Lambda$}&\multicolumn{3}{c|}{$m_g\pm\Delta m_g$}& \multicolumn{1}{c|}{R.E.} \\
\multicolumn{1}{|l|}{name}&\multicolumn{1}{c|}{(yr)}&\multicolumn{1}{c|}{$('')$}&\multicolumn{1}{c|}{(mas)}&\multicolumn{3}{c|}{(AU)}&
\multicolumn{3}{c|}{$(10^{-24}\ \mathrm{eV})$}&\multicolumn{1}{c|}{(\%)} \\
\hline
\hline
S1   &  168.4 &   48.2 & 0.22 &  369952.9 & $\pm$ & 43820.4 &  22.4 & $\pm$ & 2.7 &  11.8 \\
S2   &   16.3 &  722.1 & 0.83 &   15125.5 & $\pm$ & 884.7 & 547.9 & $\pm$ & 32.0 &   5.8 \\
S4   &   78.2 &   65.5 & 0.16 &  200418.3 & $\pm$ & 11191.1 &  41.4 & $\pm$ & 2.3 &   5.6 \\
S6   &  195.5 &  102.4 & 0.60 &  226607.6 & $\pm$ & 8807.8 &  36.6 & $\pm$ & 1.4 &   3.9 \\
S8   &   94.4 &  138.0 & 0.49 &  125957.9 & $\pm$ & 8420.6 &  65.8 & $\pm$ & 4.4 &   6.7 \\
S9   &   52.2 &  124.3 & 0.27 &  101193.1 & $\pm$ & 9289.8 &  81.9 & $\pm$ & 7.5 &   9.2 \\
S12  &   59.9 &  314.6 & 0.86 &   54047.1 & $\pm$ & 3026.3 & 153.3 & $\pm$ & 8.6 &   5.6 \\
S13  &   49.8 &   91.6 & 0.17 &  124334.4 & $\pm$ & 5855.1 &  66.7 & $\pm$ & 3.1 &   4.7 \\
S14  &   56.2 & 1465.9 & 4.02 &   16508.7 & $\pm$ & 2802.9 & 502.0 & $\pm$ & 85.2 &  17.0 \\
S17  &   77.9 &   66.1 & 0.16 &  198588.4 & $\pm$ & 16771.2 &  41.7 & $\pm$ & 3.5 &   8.4 \\
S18  &   42.6 &  107.1 & 0.18 &  102263.2 & $\pm$ & 5784.8 &  81.0 & $\pm$ & 4.6 &   5.7 \\
S19  &  137.6 &   87.1 & 0.38 &  214568.1 & $\pm$ & 89676.6 &  38.6 & $\pm$ & 16.1 &  41.8 \\
S21  &   37.6 &  217.4 & 0.41 &   56500.3 & $\pm$ & 4881.5 & 146.7 & $\pm$ & 12.7 &   8.6 \\
S22  &  550.0 &   19.0 & 0.17 & 1347095.9 & $\pm$ & 580678.1 &   6.2 & $\pm$ & 2.7 &  43.1 \\
S23  &   46.7 &  114.1 & 0.22 &  102079.9 & $\pm$ & 28448.6 &  81.2 & $\pm$ & 22.6 &  27.9 \\
S24  &  336.5 &  107.5 & 0.93 &  286723.3 & $\pm$ & 41927.1 &  28.9 & $\pm$ & 4.2 &  14.6 \\
S29  &  102.7 &   98.5 & 0.35 &  169074.0 & $\pm$ & 37807.8 &  49.0 & $\pm$ & 11.0 &  22.4 \\
S31  &  110.4 &   63.3 & 0.21 &  244347.7 & $\pm$ & 17733.9 &  33.9 & $\pm$ & 2.5 &   7.3 \\
S33  &  195.3 &   47.9 & 0.25 &  400731.7 & $\pm$ & 75407.3 &  20.7 & $\pm$ & 3.9 &  18.8 \\
S38  &   19.5 &  427.5 & 0.53 &   24533.9 & $\pm$ & 1005.0 & 337.8 & $\pm$ & 13.8 &   4.1 \\
S39  &   82.6 &  364.5 & 1.26 &   56824.5 & $\pm$ & 6638.4 & 145.8 & $\pm$ & 17.0 &  11.7 \\
S42  &  339.7 &   30.8 & 0.22 &  736342.1 & $\pm$ & 312551.0 &  11.3 & $\pm$ & 4.8 &  42.4 \\
S54  &  482.2 &   81.5 & 0.90 &  422181.5 & $\pm$ & 692183.8 &  19.6 & $\pm$ & 32.2 & 164.0 \\
S55  &   13.0 &  382.8 & 0.34 &   21721.6 & $\pm$ & 1465.5 & 381.5 & $\pm$ & 25.7 &   6.7 \\
S60  &   88.6 &  105.5 & 0.34 &  149149.2 & $\pm$ & 11131.0 &  55.6 & $\pm$ & 4.1 &   7.5 \\
S66  &  675.3 &   13.4 & 0.11 & 1933974.1 & $\pm$ & 269754.5 &   4.3 & $\pm$ & 0.6 &  13.9 \\
S67  &  438.3 &   19.3 & 0.14 & 1188222.1 & $\pm$ & 116748.7 &   7.0 & $\pm$ & 0.7 &   9.8 \\
S71  &  352.1 &  106.2 & 0.95 &  295890.2 & $\pm$ & 56006.2 &  28.0 & $\pm$ & 5.3 &  18.9 \\
S83  &  667.2 &   15.3 & 0.15 & 1737943.9 & $\pm$ & 477698.2 &   4.8 & $\pm$ & 1.3 &  27.5 \\
S85  & 3619.1 &   11.0 & 0.44 & 5195117.0 & $\pm$ & 8106789.5 &   1.6 & $\pm$ & 2.5 & 156.0 \\
S87  & 1663.8 &    7.6 & 0.12 & 4641782.4 & $\pm$ & 619016.2 &   1.8 & $\pm$ & 0.2 &  13.3 \\
S89  &  412.3 &   31.0 & 0.27 &  806547.1 & $\pm$ & 140413.3 &  10.3 & $\pm$ & 1.8 &  17.4 \\
S91  &  973.6 &   11.4 & 0.14 & 2626592.1 & $\pm$ & 322729.8 &   3.2 & $\pm$ & 0.4 &  12.3 \\
S96  &  673.2 &   13.6 & 0.12 & 1907722.2 & $\pm$ & 189196.9 &   4.3 & $\pm$ & 0.4 &   9.9 \\
S97  & 1296.3 &    9.7 & 0.15 & 3407955.4 & $\pm$ & 1361276.1 &   2.4 & $\pm$ & 1.0 &  39.9 \\
S145 &  434.8 &   23.6 & 0.19 & 1016130.7 & $\pm$ & 535791.9 &   8.2 & $\pm$ & 4.3 &  52.7 \\
S175 &   97.7 & 1812.0 & 7.23 &   18570.1 & $\pm$ & 5168.9 & 446.3 & $\pm$ & 124.2 &  27.8 \\
R34  &  893.3 &   18.6 & 0.27 & 1741793.7 & $\pm$ & 558192.6 &   4.8 & $\pm$ & 1.5 &  32.0 \\
R44  & 2825.3 &    5.5 & 0.13 & 7740489.4 & $\pm$ & 5361256.0 &   1.1 & $\pm$ & 0.7 &  69.3 \\
\hline
\end{tabular}
\label{tab2}
\end{table*}

\section{Conclusions}

A precession of orbit in Yukawa potential is in the same direction as
in GR, but dependences of precession angles on semi-major axis and
eccentricity are different in these two models, therefore, after
observations of bright star orbits for one or a few periods one could
select the best fit from two considered cases.

In paper \cite{Zakharov_16} we presented an upper bound for
graviton mass $2.9 \times 10^{-21}\ \mathrm{eV}$ using previous
observations of S2 star (see also
\cite{Zakharov_16_Quarks,Zakharov_17_Baldin,Zakharov_17_MIFI_2,
Zakharov_17_MIFI_2017,Zakharov_MIFI_18} for a more detailed discussion) and now we
also demonstrate our forecasts to reduce this upper limit. As it was
noted earlier, our current estimates for graviton mass is slightly
weaker than the LIGO ones, but it is independent and consistent with
LIGO results and we expect that the graviton mass estimate will be
significantly improved with new observations.

The obtained results show that (see Fig. \ref{fig01} and
 Table \ref{tab2}):
\begin{enumerate}
\item Range of Yukawa gravity $\Lambda$ can be constrained in  such a
way to induce the same orbital precession of stellar orbits as in GR;
\item Orbits with small eccentricities provide better constraints on
graviton mass (see Eq. (\ref{lambda_delta1_T}));
\item There is a linear dependence of $\Lambda$ constraint on
orbital periods of bright stars (perhaps, monitoring bright stars for
more than 50 -- 100 years looks rather problematic, but people
monitor comets for centuries);
\item Such a precession of stellar orbits around GC, if observed,
could provide strong constraints on the mass of graviton,
indicating that it is most likely $\approx 8 \times 10^{-23}\
\mathrm{eV}$ for orbital periods around a several decades, (see,
parameters for S9 star, for instance).
\item If we assume that the future telescopes give an opportunity to
a bright star with a period around 50 years and small eccentricity
then the similar procedure give a graviton mass constraint around
$\approx 5 - 6 \times 10^{-23}\ \mathrm{eV}$, while pericenter
advance for such a star will be $\Delta s_{GR} \approx 0.1$~mas which
is in principle should be detectable in the future.

\end{enumerate}

We expect that future observations of trajectories of bright stars
near the Galactic Center like GRAVITY \cite{blin15}, E-ELT
\cite{eelt14} and TMT \cite{tmt14} will realize our forecasts
and improve the current graviton mass constraints.

\acknowledgments

Authors thank an anonymous referee for critical remarks.
A. F. Z. thanks PIFI grant 2017VMA0014 of Chinese Academy of
Sciences, the Strategic Priority Research Program of the Chinese
Academy of Sciences (Grant No. XDB23040100), NSF (HRD-0833184) and
NASA (NNX09AV07A) at NASA CADRE and NSF CREST Centers (NCCU, Durham,
NC, USA) for a partial support. A. F. Z. thanks M. V. Sazhin for
important discussions of our previous studies on the subject and his
suggestion to find ways to improve current estimates of graviton mass
with the future observations of bright stars. P.J., D.B. and V.B.J.
wish to acknowledge the support by the Ministry of Education,
Science and Technological Development of the Republic of Serbia
through the project 176003 ''Gravitation and the large scale
structure of the Universe''.

\end{document}